# COMPARISON OF HYDROXYAPATITE AND HONEYCOMB MICRO-STRUCTURE IN BONE TISSUE ENGINEERING USING ELECTROSPUN BEADS-ON-STRING FIBERS

Nicolas Rivoallan (1,2), Marc Mueller (2), Timothée Baudequin (1), Pascale Vigneron (1), Anne Hébraud (3), Rachid Jellali (1), Quentin Dermigny (1), Anne Le Goff (1), Guy Schlatter (3), Birgit Glasmacher (2), Cécile Legallais (1)

*1 Université de technologie de Compiègne, CNRS, BMBI (Biomechanics and Bioengineering), Centre de recherche Royallieu - CS 60 319 - 60 203 Compiègne Cedex, France; 2 Institute for Multiphase Processes, Leibniz University Hannover, Hannover DE-30823, Germany; 3 ICPEES UMR 7515, Institut de Chimie et Procédés pour l'Energie, l'Environnement et la Santé, CNRS, Université de Strasbourg, 67087 Strasbourg, France.*

## Abstract

Thick honeycomb-like electrospun scaffold with nanoparticles of hydroxyapatite (nHA) recently demonstrated its potential to promote proliferation and differentiation of a murine embryonic cell line (C3H10T1/2) to osteoblasts. In order to distinguish the respective effects of the structure and the composition on cell differentiation, beads-on-string fibers were used to manufacture thick honeycomb-like scaffolds without nHA. Mechanical and biological impacts of those beads-on string fibers were evaluated. Uniaxial tensile test showed that beads-on-string fibers decreased the Young Modulus and maximal stress but kept them appropriate for tissue engineering. C3H10T1/2 were seeded and cultured for 6 days on the scaffolds without any growth factors. Viability assays revealed the biocompatibility of the beads-on-string scaffolds, with adequate cells-materials interactions observed by confocal microscopy. Alkaline phosphatase staining was performed at day 6 in order to compare the early differentiation of cells to bone fate. The measure of stained area and intensity confirmed the beneficial effect of both honeycomb structure and nHA, independently. Finally, we showed that honeycomb-like electrospun scaffolds could be relevant candidates for promoting bone fate to cells in the absence of nHA. It offers an easier and faster manufacture process, in particular in bone-interface tissue engineering, permitting to avoid the dispersion of nHA and their interaction with the other cells.

## Keywords

Tissue engineering and regenerative medicine, Bioengineering, Cell-biomaterial interactions, Biomaterials & nanotechnology, Electrospinning, Micro-structured material, Bone tissue, Beads-on-string fibers, Hydroxyapatite nanoparticles, Additive manufacturing

## 1. Introduction

Biomaterial scaffolds are one of the main factors in tissue engineering to control the cell behavior in vitro after accurate design of their composition and architecture[1]. More specifically, they can influence stem cell differentiation towards various lineages such as osteogenesis[2].

We previously developed an electrospun scaffold combining a micropatterned honeycomb structure (mimicking the native osteons[3]) and the presence of hydroxyapatite nanoparticles (nHA, main mineral component of bone) to trigger the bone differentiation of stem cells without any additional factors in the culture medium[4]. We validated the potential of this scaffold for bone tissue engineering, but we could not determine at this time whether the honeycomb structure or the calcium phosphate particles had the strongest impact on osteogenesis. nHA served two roles in the process: they provided not only their well-known osteoconductive effect[5], but they also enabled the fabrication of thicker honeycombs by allowing the creation of an electrostatic template with localized conductive paths[6]. It was therefore not possible to generate thick honeycomb-like structures without nHA to be compared to the whole composite scaffold.

However, new approaches have now been developed to produce more complex electrospun structures, and in particular beads-on-string electrospinning[7–9]. The intended generation of polymer beads along the fibers aims at replacing the nanoparticles as tools to increase the thickness of the micropatterned structures[10]. However, the presence of these beads should not negatively impact the other scaffold properties (mechanical properties, surface roughness and porosity allowing for cell adhesion, robustness, etc)[9,11].

The scientific objectives of this study were therefore (1) to confirm that the beads-on-string approaches could enable the fabrication of honeycomb-structured electrospun scaffolds without nHA, and (2) to evaluate if such scaffolds would still show mechanical properties relevant for tissue engineering. After these assumptions were proven right, we investigated (3) which parameters among honeycombs, nHA or potential synergistic effects between them had the strongest impact on osteogenesis. Embryonic murine cell line C3H10T1/2 was chosen for its ability to differentiate as bone or tendon cells. Those cells were seeded on different scaffolds without any growth factors. Different viability tests were performed, before assaying cells differentiation thanks to alkaline phosphatase (ALP) staining, which is an early marker of osteoblast differentiation.

## 2. Materials and methods

### Fabrication of the Honeycomb Collectors

Photolithography was used to manufacture honeycomb-shaped collectors for the production of micropatterned electrospun scaffolds. Briefly, SU-8 2050 (Microchem, USA) photoresist resin was deposited over a silicon wafer and spincoated at 1000 RPM during 1 min to reach a thickness of 60 µm. Honeycomb geometry of 160 µm inner-diameter and 20 µm walls (Figure 1, bottom left) was printed within the photoresist layer by UV laser lithography (Kloe Dilase, France). The photoresist layer was then developed to obtain the honeycomb micropatterns and sputter-coated with platinum during 45 seconds (Plassys MEB5505, France) in order to connect it to the ground as electrospinning collector.

### Scaffold Production by Electrospinning / Electrospraying

We developed electrospun scaffolds using either flat or honeycomb-structured collectors, resulting in random structure (R-) or honeycomb-like structure (HC-) respectively. On each collector, 3 different sets of parameters were applied, resulting in regular fibers (-R), bead-on-string fibers (-B) or regular fibers coupled to electrospraying of nHA (-HA) (Figure 1). This resulted in 6 different scaffolds: Random structure with regular fibers (R-R), random structure with beads-on-string fibers (R-B), random structure with regular fibers coupled to electrospray of nHA (R-HA), honeycomb structure with regular fibers (HC-R), honeycomb structure with beads-on-string fibers (HC-B) and honeycomb structure with regular fibers coupled to electrospray of nHA (HC-HA).

To electrospin regular and beads-on-string fibers, 10% and 12% wt of polycaprolactone (PCL, Mn = 80 kg/mol, Sigma-Aldrich, USA) were solubilized in a solvent mixture of dichloromethane (DCM, Sigma-Aldrich, USA) and N,N-dimethylformamide (DMF, Reagent Plus≥99%, Sigma-Aldrich, USA) at a ratio of 60:40 v/v for 24 h before electrospinning with a homemade set-up (voltage 17.5 kV, distance 25 cm; flow rate 1.02 mL/h; needle diameter 19G). Scaffolds with nHA were composed of a first layer of electrospun regular PCL fibers during 5 min (voltage 25 kV, distance 20 cm; flow rate 1.02 mL/h; needle diameter 19G) then it alternated with layers of electrospayed nHA and electrospun regular PCL fibers every minute for 1h. For this electrospraying of nHA, suspension of 10% wt hydroxyapatite (HA, nanopowder with a particle size ≤ 200 nm (BET), ≥97% synthetic, Sigma-Aldrich, USA,) was prepared in ethanol (Sigma-Aldrich, USA) and ultra-sonicated for 5 min (Branson Sonifier) just before electrospraying (voltage 25 kV, distance 15.5 cm; flow rate 0.6 mL/h; needle diameter 19G).

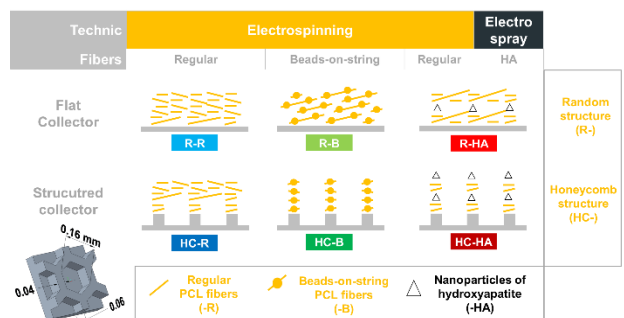

*Figure 1: Schematic representation of layer by layer 3D construction of scaffolds*

### Scanning Electron Microscopy (SEM) Characterization

Scaffold were sputter-coated with platinum (Sputter coater SC7620, EmiTech, USA) and observed by SEM (S-3400N II, Hitachi High Technologies, Japan). For observation of the thickness, scaffolds were cut in liquid nitrogen and placed in a vertical holder in the SEM. Thickness and fiber diameter were measured using imageJ software.

### Tensile Test

Scaffolds were cut into stripes (n=6) of 1x3 cm and thickness was measured with a numerical caliper (Mitutoyo Corporation, Japan). Samples were placed in metallic grips of the tensile machine (Bose Electroforce 3200, USA) and stretched at a rate of 0.1 mm/s using a cell load of 22N. The elastic modulus was calculated by analysing the stress-strain curve in the elastic zone, ending generally between 5 and 10% strain, where the relationship is linear.

### Cell Seeding on Scaffolds

The embryonic murine cell line C3H10T1/2 (ATCC CL-226) was cultured on Corning T-75 flasks at 85% confluence with DMEM (Hyclone, USA) supplemented with 10% fetal bovine serum (FBS, Gibco Invitrogen, USA), 1% of glutamine (Gibco Invitrogen, USA), and 1% of penicillin–streptomycin (Gibco Invitrogen, USA) at 37°C, 5% $CO_2$. The scaffolds were cut into discs of 0.6 cm diameter with biopsy punch, disinfected with ethanol 70% (Sigma-Aldrich, USA) for 45 min, washed 3 times with PBS 7.4 (phosphate buffered saline, Gibco Invitrogen, USA) and incubated in medium overnight before the cell culture experiment. Cells were seeded on the scaffolds at a density of 100 000 cells/cm² and incubated 1 h before adding culture medium. Scaffolds were placed in new well-plates before each analysis in order to avoid the influence of residual cells in the bottom wells. The complete medium was changed every 2 days.

### Cell Viability and Proliferation

In order to verify the effect of the beads-on-string fibrous scaffold on cell culture, cell viability and proliferation were compared among R-R, HC-B and HC-HA scaffolds. Cell proliferation was evaluated using MTS assay (CellTiter 96® AQueous One Solution

Cell Proliferation Assay (MTS), Promega, Germany) at 6 days of culture and Alamar Blue (Sigma-Aldrich, USA) (10% solution in complete medium) (Molecular Probes) after 2h of incubation at 2, 4 and 6 days of culture with R-R scaffold without cells as control. For MTS, 490 nm was used for measuring absorbance while for Alamar Blue, 535 nm and 590 nm wavelength were used for excitation and emission fluorescence respectively (TECAN, Switzerland)

After 6 days of culture on the scaffolds, cell viability was estimated with a Live/Dead kit (Invitrogen, USA). Calcein AM (1 mM) and ethidium homodimer-1 (1 mM) fluorescent dyes were used to stain viable and dead cells, respectively. The samples were observed using fluorescence microscopy (Leica Microsystem, Germany), allowing us to qualitatively determine cell viability and distribution.

Moreover, cells were observed at different timepoints with Inverted confocal microscope (Zeiss 710, Germany). Samples were fixed with 4% (w/v) paraformaldehyde solution (Agar Scientific, United Kingdom) in PBS for 10 min then permeabilized with 0.5% Triton X-100 (VWR, United Kingdom) for 10 min. Nonspecific binding sites were blocked by incubating the substrates in 1% (w/v) BSA (Sigma-Aldrich, USA) in PBS for 15 min. The staining solution rhodamine phalloidin (Invitrogen, USA) to selectively stain the F-actin was added at 5 U/mL for 45 min. Samples were then washed in PBS and mounted between glass slice in Mowiol Mounting Medium (Sigma-Aldrich, USA).

### Alkaline Phosphatase (ALP) staining

After 6 days of culture, complete medium was replaced with BCIP/NBT (Sigma-Aldrich, USA) solution to stain Alkaline Phosphatase and incubated overnight at room temperature without light. Area and saturation of the ALP staining pictures were measured using imageJ. Briefly, pictures were inverted and converted to Hue, Saturation, Brightness (HSB) stack to obtain the inverted brightness pictures and respective histograms. The same threshold was applied to all pictures to remove pixels corresponding to the scaffold without staining. For area of ALP staining, number of pixels after threshold was divided by total pixel number. For intensity of ALP staining, inverted brightness (0 for white, 1 for black) was multiplied by the corresponding number of pixels before dividing by the total pixel number post-thresholding (Figure 2).

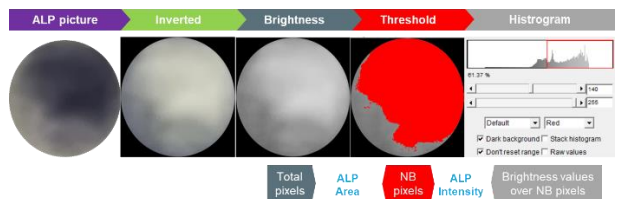

*Figure 2 : Image processing for ALP staining area and intensity evaluation*

### Statistical Analysis

All data are represented as mean ± standard deviation (SD). ANOVA statistical test was used to define the significance of the results (p<0.05), Mann-Whitney was used for comparing two populations while Tukey parametric test was used for comparing more than two populations.

## Results

### Production of thick honeycomb structure

Honeycomb structure was obtained by electrospinning PCL on honeycomb-structured wafer as collector. Beads-on-string fibers permitted to increase the thickness of the honeycomb structure compared to classical fibers. Thickness of the honeycomb structure reached 60 µm with beads-on-string fibers and only 20 µm with commonly used smooth fibers (Figure 3). The honeycombs did not appear as clearly defined with beads, but still offered an inner diameter of 100 µm and walls of 40 µm.

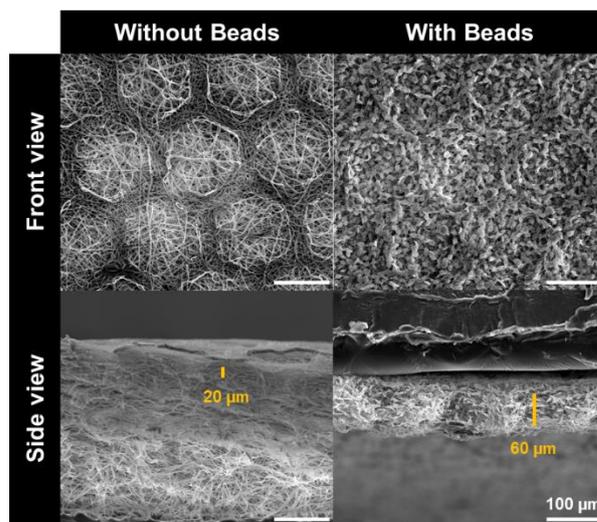

*Figure 3 : SEM pictures of honeycomb structured scaffold with and without beads-on-string fibers (scale = 100 µm).*

### Mechanical and Biological impacts of beads-on-string fibers

Beads-on-string fibers decreased the mechanical properties of the scaffold (Table 1) compared to classical fibers. Significant decrease (p=0.0087) of the Young modulus was observed from 12.46 MPa to 8.74 MPa for fibers without and with beads respectively in randomly structured scaffolds (R-R, R-B). This could be explained by the reduction observed in fiber diameter from 1.34 µm to 0.34 µm when beads were present.

| Properties | Without beads | With beads | p value |
|---|---|---|---|
| Young Modulus (MPa) | 12.46 ± 1.12 | 8.74 ± 0.66 | 0.0087 |
| Maximal stress (MPa) | 4.44 ± 1.06 | 3.86 ± 1.35 | 0.0931 |
| Fiber diameter (µm) | 1.34 ± 0.07 | 0.34 ± 0.03 | < 0.0001 |

*Table 1: Mechanical properties of random structure scaffolds with different kind of fibers.*

In order to evaluate the viability of cells on beads-on-string fibers before comparing osteogenic promotion, the scaffolds were seeded with C3H10T1/2 cells at a density of 100 000 cells/cm$^2$ in classical culture conditions without any growth or differentiation factors. Alamar blue tests were performed at day 2, 4 and 6 to evaluate the metabolic activity. Although, at day 6, the difference was significant between HC-HA and R-R only, these results showed that the presence of beads did not jeopardize cell growth (Figure 4 A). To confirm that, MTS assay was performed at day 6 too and shown close similar tendency (data not shown). Finally, after 6 days of culture, Live/Dead assays were performed to observe the cell viability on the scaffolds. They showed excellent cell viability with most of living cells in contact with both types of fibers of the scaffolds (Figure 4B). Confocal microscopy observations of immunostaining of the nucleus (blue) and cytoskeleton (green) confirmed the spreading of the cells on the biomaterial (Figure 4C).

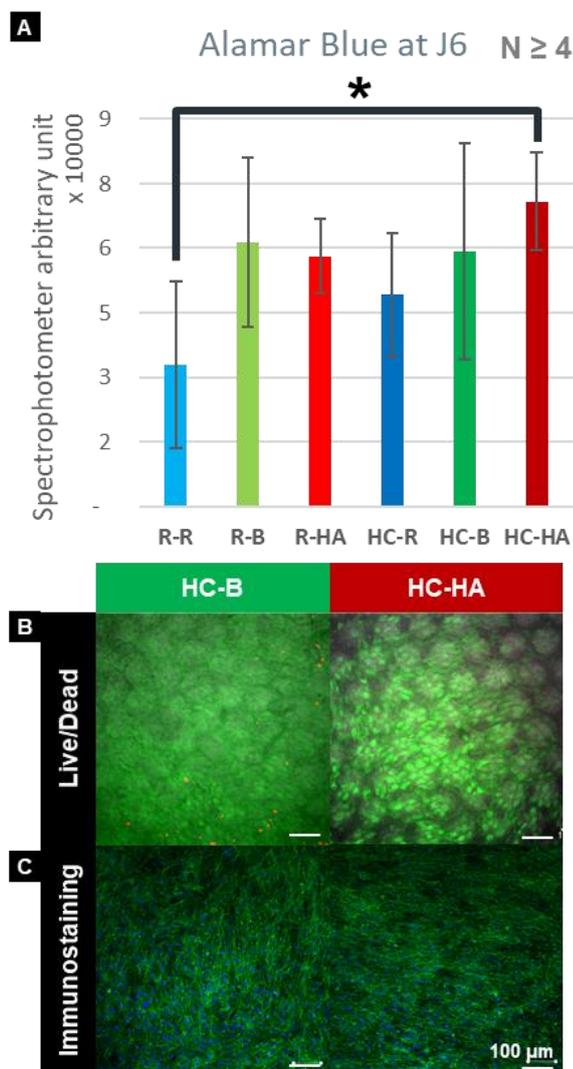

*Figure 4 : (A) Metabolic activity of cells after 6 days on different scaffolds. (B) Epifluorescence pictures of Live/Dead and (C) Confocal pictures of Honeycomb scaffolds composed of beads-on-string fibers (on the right) and layers of HA respectively (on the left). (means, standard deviation; scale = 100 µm; n≥4)*

### Cell differentiation

The early bone differentiation of C3H10T1/2 was evaluated thanks to quantification of area and intensity of ALP staining (Figure 5). We can distinguish two distinct groups. The first is composed of scaffolds composed of honeycomb-like structure and/or presence of nHA (HC-B, HC-R, HC-HA, R-HA) with high area and intensity of ALP staining while the second group shown smaller area and much lower intensity (R-R, R-B).

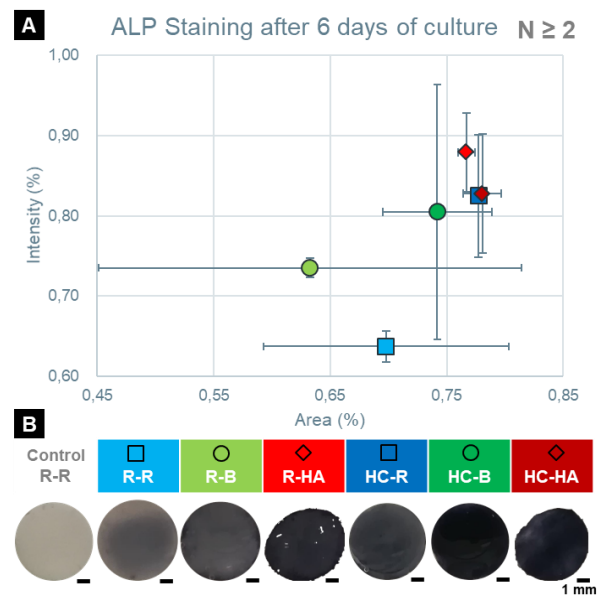

*Figure 5 : (A) Measurement of area and intensity of the staining thanks to imageJ. (B) Picture of the scaffold after incubation in ALP staining overnight at day 6. (means, standard deviation; scale = 1 mm; n≥2)*

## 3. Discussion

In a previous work, we demonstrated that the honeycomb structure composed of PCL electrospun fibers and nHA was a relevant candidate for a semi-3D organized support for bone regeneration[4]. The nHA, in addition to their well-known beneficial effect on osteosynthesis, were necessary to produce thick structure. When electroprayed, they interact with the electrospinning collector to better follow its geometry, contrary to classical electrospinning process. Alternate layers of PCL electrospinning and nHA electrospraying permitted therefore to reach thickness of 50 µm, which was considered as semi-3D. It was however impossible to achieve comparable thicknesses when removing nHA from the process, as the interactions with the structured collector decreased gradually with the accumulation of fiber layers.

Beads-on-string fibers can be produced during unstable electrospinning in contrast to regular fibers[12]. Those fibers are composed of beads linked to each other by fibers with smaller diameter. Those kind of fiber structures are usually avoided because it was considered an undesired defect[13,14]. Indeed, it reduces the uniformity of nanofibers, and so, offers lower mechanical properties compared to regular ones as well as poor repeatability of the process. This reduction is mainly caused by the decrease in fiber diameter[15,16]. Nevertheless, beads-on-string fibers seemed to get interest for drug release in tissue engineering[17]. Here, those beads-on-string permitted to increase the thickness of honeycomb-like structure without nanoparticles, as hypothesized following the approach described by Nedjari et al[10]. Briefly, fibers composed of two kinds of diameters (beads diameters and fibers-between-beads diameter) were capable of self-assembly. The use of a microstructured collector allowed the control of the dimensions of this self-assembled structure.

Using beads-on string fibers made therefore possible to produce thick honeycomb-like scaffold without nanoparticles using the same strategy. In order to evaluate whether osteogenic effect came from the structure (HC) of from the composition (presence of nHA), beads-on-string fibers were used to produce successfully thick honeycomb structure without nHA. The scaffold reached a thickness of 60 µm, comparable to scaffolds using nHA[4].

Uniaxial tensile test highlighted a decrease of the Young Modulus from 12.46 MPa to 8.74 MPa compared to regular fibers. Nevertheless, those mechanical properties were in the same order of magnitude (young modulus and maximal stress), still allowing easy manipulation for experiments or even future translational steps[18].

Then, C3H10T1/2 cells were chosen for their ability to differentiate as bone or tendon cells, and seeded on different scaffolds to evaluate their cytotoxicity. Scaffolds composed first of regular fibers with random structure (R-R), second of beads-on-string fibers with honeycomb like structure (HC-B) and last of regular fibers coupled with layer of nHA with honeycomb structure (HC-HA); were seeded with C3H10T1/2 at a density of 100000 cells/cm². After 6 days of culture without any growth factors, Alamar Blue, MTS and Live/Dead tests were performed and did not show any excessive mortality induced by beads-on-string fibers. Immunostaining of nucleus and cytoskeleton observed with confocal microscopy confirmed the correct colonization in the scaffolds.

Finally, ALP staining was performed on scaffold after 6 days of culture without any growth factor again. ALP staining analysis showed a clear response of R-HA and HC-HA on area and intensity of the staining, confirming the relevant role of nHA to stimulate osteogenesis[19–21]. However, honeycomb structure scaffold (HC-R, HC-B) permitted to reach similar effect without nHA, if compared to a random PCL scaffold, confirming the topographic value of the honeycomb structure. The highest level of ALP production was reached thanks to the scaffolds combining nHA and HC, highlighting that both effects can still act together to speed up the stem cell differentiation. This will need to be confirmed with future studies investigating more closely osteogenesis with late markers and multiple analyses, but this first differentiation study showed promising preliminary results.

In addition, beads-on-string fibers can generate a larger surface area[13,22]. Taken together, these results showed that thick honeycomb-structured scaffolds based on beads-on-string could be an interesting approach to propose an easier and faster production process, compared to alternating electrospinning of PCL and electrospraying of nHA. This method is indeed time consuming and requires complex set-up, when the approach presented here used only a standard electrospinning device. As an alternative, many studies proposed also to incorporate nHA directly in the polymer solution for electrospinning[23–26]. Nonetheless, electrospinning of nHA leads generally to particles repartition inside the fibers, increasing the mechanical properties but limiting the interactions between the cells and those bioceramics. In another study, Ramier et al[27] proposed to electrospray nHA and electrospin polymer simultaneously in order to spread the nHA onto the fibers. In the same purpose, different electrospun scaffolds were soaked in calcium solution and then in simulated body fluid solution to coat the scaffold with apatite[28]. However, this process of double soaking was time-consuming and could last 24h and one week respectively[29], indicating again a process longer than only electrospinning of beads-on-string fibers with honeycomb structure. It could lead to promising application for bone-interfaces tissue engineering by avoiding the risk of release and contact of nHA with cells which are not expected to differentiate into bone tissue (tendon, ligament, cartilage…).

## 4. Conclusion

Beads-on-string fibers permitted to increase the thickness of honeycomb-structured electrospun scaffolds without using nHA. The use of those beads in the scaffold did not alter the cell behavior but decreased the mechanical properties. However, those mechanical properties still allowed easy manipulation for tissue engineering and relevant range of elastic modulus. It was then possible to evaluate the role of the structure on the differentiation of stem cells towards the bone lineage compared to the effect of the nHA themselves. We were therefore able to show the positive effects of the structure as well as the nHA. Both composition and structure are interesting for bone tissue engineering, and can also act together to reach highest levels of early differentiation. However, using only structure enables an easier and faster manufacture of the scaffold, moreover avoiding the risk of nHA dispersion in the medium, especially in a case of bone-interfaces tissue engineering where those particles can interact with other cells that are not supposed to follow bone fate.

## Acknowledgements

This work was funded by the Labex MS2T (Challenge Interfaces) supported by the French Government, through the program "Investments for the Future" managed by the National Agency for Research (Reference ANR-11-IDEX-0004-02), by the project TENORS (Reference ANR-21-CE18-0035-01). NR received a scholarship from Graduierten Akademie from Leibniz University of Hannover, fundings from the French Ministry of Higher Education and Research, and was supported by German Academic Exchange Service (DAAD - Funding 57681230).